\documentclass[9pt,twocolumn,twoside]{pnas-new}
\templatetype{pnasresearcharticle} % Choose template 
\usepackage{booktabs,threeparttable}
\title{Copula-based Risk Aggregation with Trapped Ion Quantum Computers}

\author[a,1,2]{Daiwei Zhu}
\author[b,1]{Weiwei Shen} 
\author[b]{Annarita Giani}
\author[b]{Saikat Ray Majumder}
\author[b]{Bogdan Neculaes} 
\author[a]{Sonika Johri}

\affil[a]{IonQ Inc., 4505 Campus Drive, College Park, Maryland
}
\affil[b]{GE Research, One Research Circle, Niskayuna, New York
}

\leadauthor{Zhu} 

\significancestatement{Risk management is at the core of business decision-making in different industries such as finance, supply chain and manufacturing. Assessing risk requires modeling interdependencies among random variables. Recent insights suggest that quantum computers can efficiently model such interdependencies by leveraging quantum entanglement among qubits. To study the viability of the quantum approach to real-world applications, we formulate it to model up to 4 real-world indices in the stock market. Our quantum framework outperforms its classical counterpart in some tests while facing scaling-up challenges. To address such challenges, we introduce a novel optimization technique that potentially impacts quantum algorithm research well beyond the risk management topic studied here.
}

\authorcontributions{D.Z., W.S., A.G., S.R.M, B.N. and S.J. designed research. W.S., A.G., S.R.M, and B.N. acquired and prepared the stock market data for the study. W.S., A.G., S.R.M, and B.N. implemented the classical model. D.Z. and S.j. formulated the quantum model. D.Z. and W.S. performed numerical simulation of the quantum model. D.Z. and S.J. performed experiments on the quantum computer. D.Z. and W.S. analyzed the experimental data. All authors contributed to the preparation of the manuscript.}

\authordeclaration{The authors declare no competing interest.}

\equalauthors{\textsuperscript{1}D.Z. and W.S. contributed equally to this work.}

\correspondingauthor{\textsuperscript{2}Correspondence should be addressed to \href{mailto:zhu@ionq.co}{zhu@ionq.co} }

\keywords{Quantum Computing $|$ Ion Trap $|$ Copula $|$ Risk Aggregation} 

\begin{abstract}
Copulas are mathematical tools for modeling joint probability distributions. Since copulas enable one to conveniently treat the marginal distribution of each variable and the interdependencies among variables separately, in the past 60 years they have become an essential analysis tool on classical computers in various fields ranging from quantitative finance and civil engineering to signal processing and medicine. The recent finding that copulas can be expressed as maximally entangled quantum states \cite{zhu2021generative} has revealed a promising approach to practical quantum advantages: performing tasks faster, requiring less memory, or, as we show, yielding better predictions. Studying the scalability of this quantum approach as both the precision and the number of modeled variables increase is crucial for its adoption in real-world applications. In this paper, we successfully apply a Quantum Circuit Born Machine (QCBM) based approach to modeling 3- and 4-variable copulas on trapped ion quantum computers. We study the training of QCBMs with different levels of precision and circuit design on a simulator and a state-of-the-art trapped ion quantum computer. We observe decreased training efficacy due to the increased complexity in parameter optimization as the models scale up. To address this challenge, we introduce an annealing-inspired strategy that dramatically improves the training results. In our end-to-end tests, various configurations of the quantum models make a comparable or better prediction in risk aggregation tasks than the standard classical models. Our detailed study of the copula paradigm using quantum computing opens opportunities for its deployment in various industries.
\end{abstract}

\dates{This manuscript was compiled on \today}
\doi{\url{www.pnas.org/cgi/doi/10.1073/pnas.XXXXXXXXXX}}

\begin{document}

\maketitle
\ifthenelse{\boolean{shortarticle}}{\ifthenelse{\boolean{singlecolumn}}{\abscontentformatted}{\abscontent}}{}

% If your first paragraph (i.e. with the \dropcap) contains a list environment (quote, quotation, theorem, definition, enumerate, itemize...), the line after the list may have some extra indentation. If this is the case, add \parshape=0 to the end of the list environment.
\dropcap{J}oint modeling of several random variables is required in analyzing multidimensional events, from assessing climate change, to predicting economic cycles, from identifying the cause of illnesses to guarding against catastrophic events and cyberattacks. Linear correlation, because of its simplicity in calculation and its equivalence to dependence when variables follow the normal distribution, has been widely adopted in data-based analysis and decision-making for multiple variables.  However, its sole measure of linear relationship between two variables sets limitations and even pitfalls, particularly when the true distribution differs considerably from the normal distribution \citep{EmbrechtsMcNeilStraumann2002}. Meanwhile, the concept of a copula that expresses dependence on a quantile scale oﬀers a richer representation of dependence beyond linearity and normality \citep{Sklar1959FonctionsDR}, with applications in finance, engineering and medicine \citep{joe2014dependence,cherubini2004copula,lebrun2009innovating,lambert2002copula}. 

In risk management, the dependence concept is crucial because it formalizes the idea of undiversifiable risks. Financial institutions then determine risk-based capital reserve accordingly. Undiversifiable risk, also called aggregate risk or systematic risk, refers to the vulnerability to factors that impact the outcomes for various aggregate financial vehicles, such as the broad stock market. While pairwise dependence is often captured by joint distributions, it imposes a tight constraint that the marginal distribution in each dimension should be in the same family as the associated joint distribution \citep{Genest2007EverythingYA}. As they allow for modeling the joint distribution of a random vector by estimating its marginal distributions and dependence structure separately, copulas facilitate a practically desired approach of building multivariate risk models, where the marginal behavior of individual risk factors are often known better than their dependence structure \citep{mcneil2010quantitative}. In practice, as financial institutions are exposed to multiple types of risk, risk management and aggregation via dependence structure on the corporate level are required by both daily operation and regulation \citep{goodhart2022holistic}. In credit risk modeling, risk factors are selected and aggregated by Gaussian copulas in the GCorr risk model by Moody's Analytics \citep{Moody2012}. For derivatives pricing, the risk profile and pricing of collateralized debt obligations (CDOs) are mainly based on Gaussian copulas \citep{li2000default}. In addition, the recently developed vine copulas enable the flexible modelling of the dependence structure for portfolios in high dimensions \citep{aas2009pair}.

%{\color{red}Can we talk about the challenges of modeling %arbitrary/empirical copulas in higher dimensions here? Any ML %approaches?}
While various bivariate copulas exist, how to construct copulas in higher dimensions is less clear. The three main classes of copulas, namely, Archimedean, vine and elliptical copulas, have their inherent shortcomings. Archimedean copulas lack flexibility in high dimensions due to their limited parameters. Vine copulas that offer greater flexibility by increasing complexity in the modeling process by decomposing the density into conditional bivariate densities are prone to overfitting in applications. For example, in the modeling process for a ten dimensional canonical vine copulas, over one million decompositions are needed \citep{mazo2015class}. Elliptical copulas assume undesired dependence structure such as similar tail symmetry among all pairs of variables \cite{durante2010construction}. As pointed out by the survey paper \cite{elidan2013copulas}, limited algorithmic innovation in the copula community is borrowed from machine learning for automatically inferring the model structure from observed data. With the imposed mathematical formulations, the flexibility of the model is restricted by the assumed parameterization. In contrast, we present a data driven quantum method in this paper which does not make assumptions about the parametric forms of the dependence structure, and thus has a higher degree of modeling flexibility. %Thus, we believe that the synergy of data driven methods and copula modeling on quantum computers is a worthwhile exploration to accurately represent the structure of the data.   

%Pairwise dependence is often captured using classical bivariate %distributions. But in this case the two individual variables must %have the same family of univariate %distributions~\cite{Genest2007EverythingYA}. Copula is an approach %to modeling dependence between variables based on the Sklar %theorem~\cite{Sklar1959FonctionsDR}.
%\begin{itemize}
%\item Problem Description
%\item Add background
%\end{itemize}
%Previous/related work (Adding results of IonQ and fidelity paper, %what do we add here.)

Following the machine learning approaches to copulas, it has been demonstrated that a generative learning algorithm on trapped ion quantum computers for up to 8 qubits outperformed equivalent classical generative learning models with the same number of parameters in terms of the Kolmogorov–Smirnov (KS) test ~\cite{zhu2021generative}. In that work, a Quantum Generative Adversarial Network (QGAN) and a Quantum Circuit Born Machine (QCBM) were trained to generate samples from joint distributions of historical returns of two individual stocks from the technology sector. While this work outlined a general quantum implementation of copulas, numerical results from a quantum simulator and from quantum hardware were restricted to two variables. The scalability of the technique, which is at the core of the potential quantum advantage offered by this approach, is yet to be comprehensively analyzed. An end-to-end evaluation of how such an approach would perform in actual applications is also not developed. To explore these directions, in this work, we increase the number of variables to three and four. We perform training on the latest generation quantum computer from IonQ (\textit{IonQ Aria}) with up to 8 qubits, and evaluate the trained model with up to 16 qubits. Our quantum approach outperforms the classical methods in some of the end-to-end tests. 
We also observe a drop in the training efficacy as the model becomes larger or uses more qubits to support higher precision. To address this issue, we develop specialized techniques to train the parametric quantum circuits in higher dimensions that can be applied to hybrid quantum algorithms beyond the application domains in this paper. Finally, in the supplementary materials, we present a workflow to perform an end-to-end test that evaluates the efficacy of the model in real-world risk aggregation tests. 
%We demonstrate the first quantum copula scaling to 3 and 4 variables and we show that the quantum approach outperforms the classical methods in some cases. These results set the foundation for potential deployment of this vision towards real life applications in multiple industries. [I would recommend against repeating these sentences since they already appeared in abstract and conclusion.]

\section*{Problem Description}
In our study, we model the returns of four representative stock market indices - DJI, VIX, N225 and RUT - by copulas. 
\begin{enumerate}
  \item {\bf Dow Jones Industrial Average (DJI)} is the average of stock prices of 30 selected large and influential U.S. companies. 
  \item {\bf Market Volatility Index (VIX)} created by the Chicago Board Options Exchange measures the stock market volatility through options on the S\&P 500 Index.
  \item {\bf Japan Nikkei Market Index (N225)} is a stock market index including Japan’s top 225 companies that represent Japanese economy after the World War II.
  \item {\bf Russell 2000 Index (RUT)} is a stock market index that tracks the performance of 2000 small-cap companies in the U.S.
\end{enumerate}
\begin{table}[t]
\begin{tabular}{ c c c c c c c c} 
\toprule
$m$ & Assets & $\mu$ $(\%)$ & $\sigma$ $(\%)$ & $s$ & $\kappa$ & Min ($\%$) & Max ($\%$)\\
\midrule
 4729 & DJI & 0.02 & 1.2 & -0.5 & 16.5 & -13.8 & 10.8\\ 
4729 & VIX & -0.02 & 7.1 & 1.1 & 9.9 & -35.1 & 76.8\\ 
4729 & N225 & 0.01 & 1.5 & -0.4 & 9.5 & -12.1 & 13.2\\ 
4729 & RUT & 0.02 & 1.6 & -0.6 & 10.5 & -15.4 & 9.0\\ 
\bottomrule
\end{tabular}
\caption{Asset return statistics before standardization. $m$ is the number of data points, $\mu$ is the mean of the return, $\sigma$ represents the standard deviation, $s$ is the skewness and $\kappa$ is the kurtosis. Min and Max represent the minimum and the maximum values, respectively.}
\label{tab:data} 
\vspace{-12pt}
\end{table}

%\begin{figure}[ht]
%\centering
% \includegraphics[width=0.45\textwidth]{graphics/4assets.png}
% \caption{Time series representation of the four indexes analysed.}
% \label{fig:data_plots}
%\end{figure}

\begin{figure}[t]
\centering
	\includegraphics[width=0.4\textwidth]{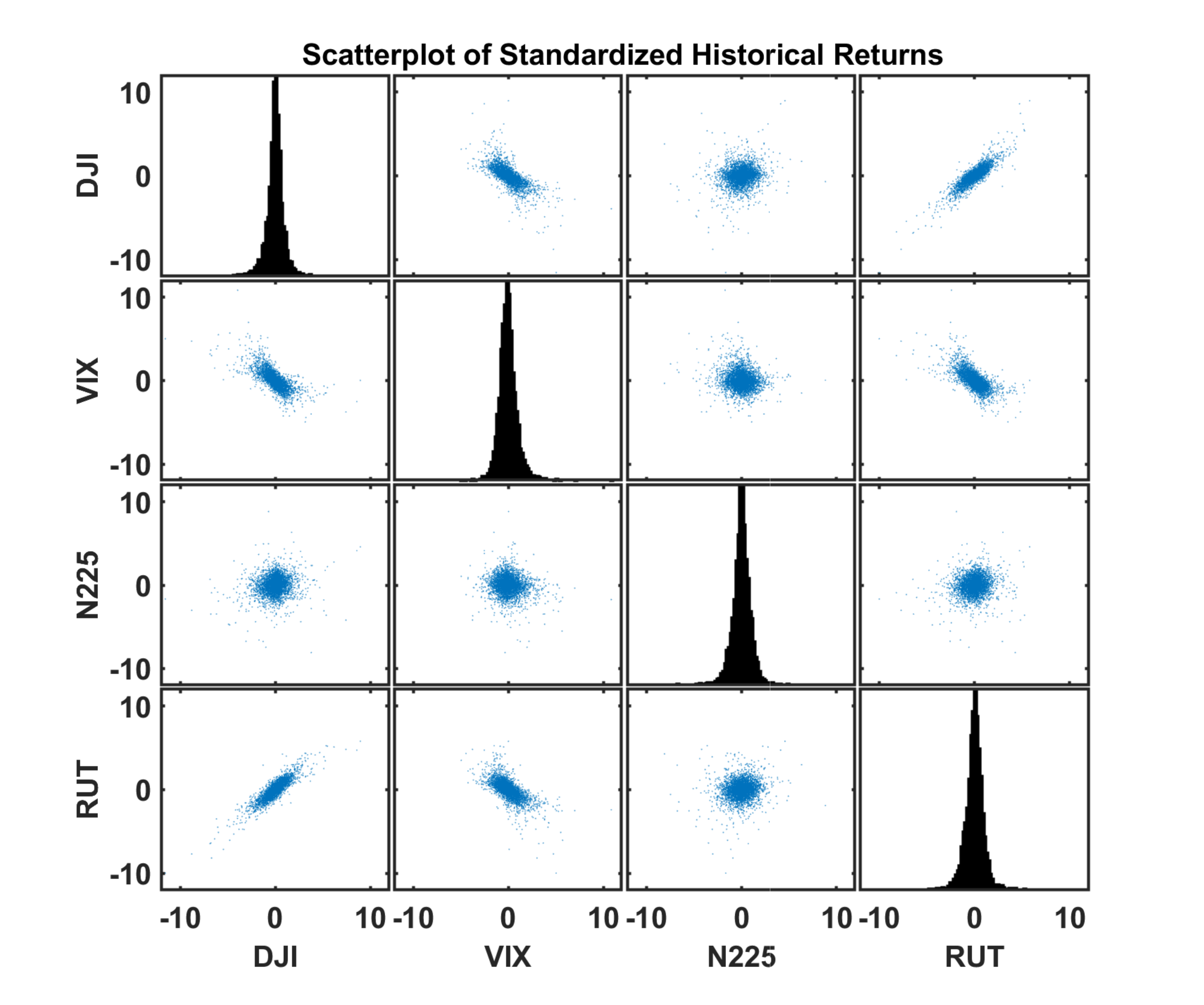}
 \caption{Scatterplot of the standardized returns of four indices in Tab. \ref{tab:data}. The diagonal subplots are the histograms for the four assets. Among those pairwise subplots, \text{DJI} and \text{RUT} are highly positively correlated. \text{DJI} and \text{RUT} are both negatively correlated with \text{VIX}. \text{N225} has mild dependence with the U.S. market and \text{VIX}.}
 \label{fig:scatterdata_plots}
\end{figure}

They are easily accessible and have long periods of time from 01/04/2001 to 12/30/2020. These years reflect the vicissitudes of the market environment, such as multiple financial crashes and booms. On the one hand, \text{DJI} and \text{RUT} highlight the long-term performance of the U.S. market with their limited selection bias. They are highly positively correlated. On the other hand, \text{N225} represents a foreign index with mild dependence with the U.S. market and \text{VIX}, as a market fear gauge, commonly negatively correlates with market indices. Thus, through empirical evaluations on those data, we can thoroughly understand the performance of different approaches in various domains in the dependence spectrum. After the data cleaning step each index has 4729 daily log returns. The data with computed statistics are shown in Tab.~\ref{tab:data}. In our study, we assume returns of each index are independent and identically distributed in time and standardize the four indices. Fig.~\ref{fig:scatterdata_plots} illustrates the diverse dependence relations between indices as discussed. This study models the joint distribution of the returns of the four assets via copulas and then tests the corresponding risk estimates for an equally-weighted portfolio of those four assets, where both classical and quantum methods are employed in the modeling step for comparison.

\section*{Classical Approach}
In this section, we first recapitulate the copula framework used in this paper. Then we describe the steps of generating new samples from the given datasets by the classical approach.  Illustration over a wide range of applications and theory of copulas may be found in the monograph \cite{joe2014dependence}. 

Suppose that a random variable $X$ has a continuous distribution. Then the random variable $U_X=F_X(X)$ follows a standard uniform distribution, where $F_X(\cdot)$ is its cumulative distribution function (CDF).  This process is called the probability integral transform. It underlies the copula approach which is the transformation of a joint distribution into a set of marginal distributions used with a dependence function called a copula $C(\cdot)$. The copula $C$ is a multivariate distribution function with marginals following standard uniform distributions.

Sklar's theorem states that if $F(x_1,\dots,x_n)$ is a joint distribution function with marginal distributions $F_1 (x_1 ),\dots,F_n (x_n)$, then there exists a copula function $C:[0,1]^n\rightarrow[0,1]$ such that
\begin{equation}\label{eq:sklar}
	F(x_1,\dots,x_n)=C(F_1(x_1),\dots,F_n(x_n)),
\end{equation} 
or
\begin{equation}\label{eq:sklar2}
	C(u_1,\dots,u_n) = F(F_1^{-1}(u_1),\dots,F_n^{-1}(u_n)),
\end{equation} 
where we call $(x_1,\dots,x_n)\in \mathbb{R}^n$ as the variable in the original space and $(u_1,\dots,u_n) \in [0,1]^n$ as the variable in the copula space. \eqref{eq:sklar} offers the steps of building a copula model from input $(x_1,\dots,x_n)$ in the original space, while \eqref{eq:sklar2} shows how we use a built copula model from the transformed input $(u_1,\dots,u_n)$ in the copula space. We provide a constructive algorithm combining the two steps starting from data inputs. Note that the sample in the copula space will be called the pseudo-sample. Denote by $\mathbf{X}\in \mathbb{R}^{M\times n}$ the input data matrix, $\mathbf{Y}\in \mathbb{R}^{N\times n}$ the random sample matrix, and $\mathbf{U}_X\in \mathbb{R}^{N\times n}$ the random pseudo-sample matrix after applying probability integral transform into each column of the input data matrix $\mathbf{X}$, where $M$ is the number of $n$-dimensional input data and $N$ is the desired number of simulated samples. Denote by $\mathbf{x}_{i\cdot}$ the $i$-th row of the data matrix $\mathbf{X}$, $\mathbf{x}_{\cdot j}$ the $j$-th column of input data matrix, and ${x}_{ij}$ the element in the $i$-th row and $j$-th column of the data matrix. Denote by $\mathbf{u}_{i\cdot}$ the $i$-th row of the random pseudo-sample matrix, and $\mathbf{u}_{\cdot j}$ the $j$-th column of random pseudo-sample matrix. Denote by $\mathbf{y}_{i\cdot}$ the $i$-th row of the random sample matrix, and $\mathbf{y}_{\cdot j}$ the $j$-th column of random sample matrix. Alg.~\ref{alg: SSRPConstruction} summarizes the steps of simulating $N$ $n$-dimensional random sample points $\mathbf{Y}$ from $M$ $n$-dimensional input data points $\mathbf{X}$. 
\begin{algorithm}[t]%\fontsize{8.5pt}{9.5pt}\selectfont
	\caption{Classical Copula Modeling}
	\begin{algorithmic}[1]
		\State {\bfseries Inputs:} Data matrix $\mathbf{X}$, size of desired sample $N$;
		\State Estimate the marginal distribution $\hat{{F}}_j$ from each column $\mathbf{x}_{\cdot j}$, $j=1,\dots,n$;
		\State Compute the probability integral transform $\hat{{F}}_j(\mathbf{x}_{\cdot j})$, $j=1,\dots,n$;
		\State Select and calibrate a copula $\hat{C}$ by the maximum likelihood method according to the pseudo-samples $\hat{F}_j(x_{ij})$, $i=1,\dots,M$ and $j=1,\dots,n$;
		\State Generate $N$ $n$-dimensional random numbers $\mathbf{u}_{i\cdot}$ from the copula $\hat{C}$, $i=1,\dots,N$;
		\State Generate $N$ $n$-dimensional random numbers $\mathbf{y}_{\cdot j}$ by transforming $\hat{{F}_j}^{-1}(\mathbf{u}_{\cdot j})$, $j=1,\dots,n$;
		\State {\bfseries Outputs:} Random sample $\mathbf{Y}$.
	\end{algorithmic}
	\label{alg: SSRPConstruction}
\end{algorithm}
\begin{figure}[t]
\centering
	\includegraphics[width=0.4\textwidth]{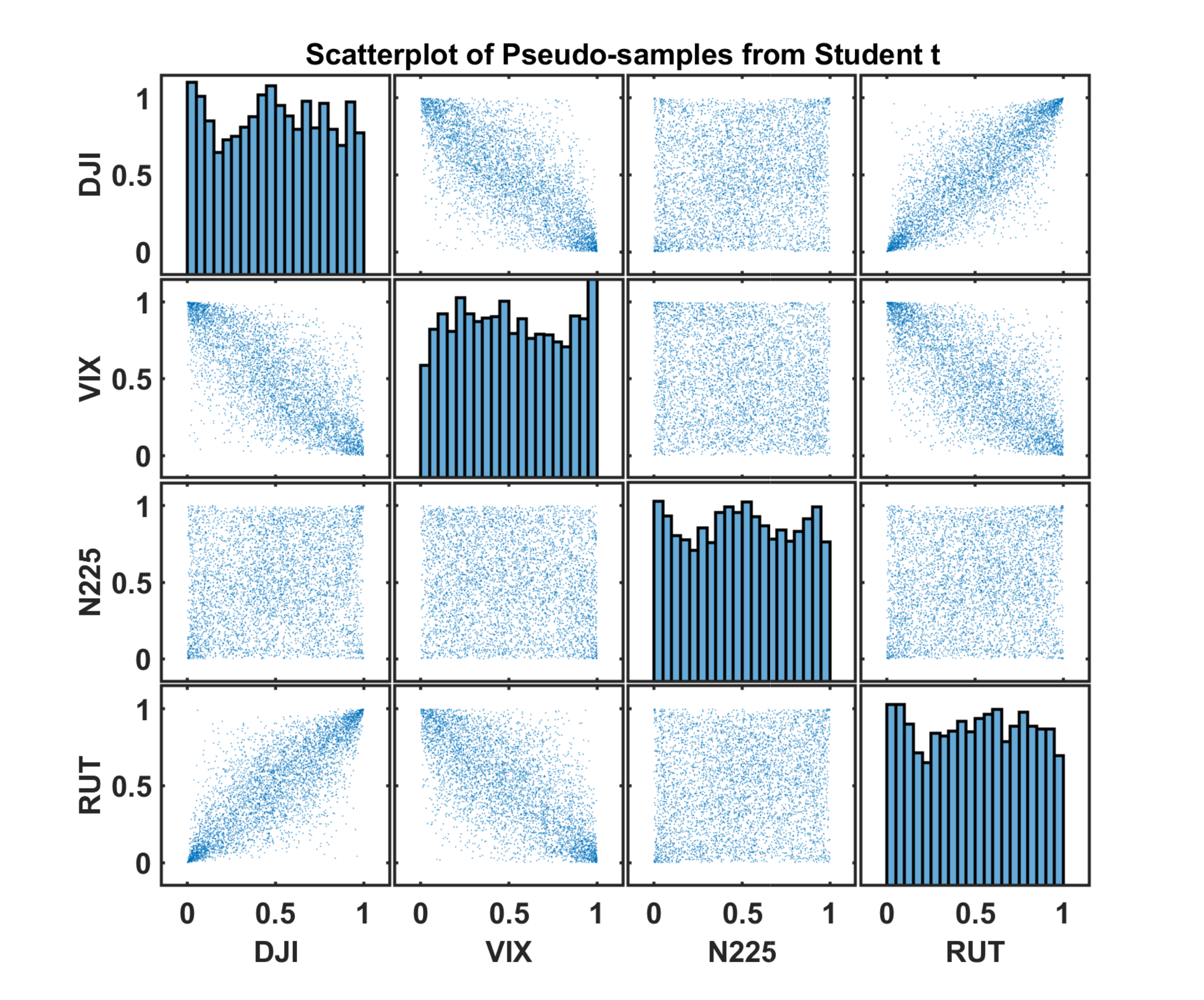}
 \caption{Pseudo-samples in the copula space of the four indices in Tab. \ref{tab:data}. The diagonal subplots are the histograms of the pseudo-samples of the four indices. Given the finite number of data points, the histograms are not perfect standard uniform distributions.  Among those pairwise subplots, \text{DJI} and \text{RUT} are highly positively correlated. \text{DJI} and \text{RUT} are both negatively correlated with \text{VIX}. \text{N225} has mild dependence with the U.S. market and \text{VIX}. Compared with Fig. \ref{fig:scatterdata_plots}, the dependence structure has been more clearly revealed. }
 \label{fig:ionq_psedu_sample_T_four}
\end{figure}

We apply the above algorithm to model the returns of the four indices.  In line with known stylized results in equity returns, Student's t-distribution is used for the marginal distributions and t-copula is chosen for the dependence structure \citep{cont2001empirical, zeevi2002beyond}. Fig.~\ref{fig:ionq_psedu_sample_T_four} illustrates the pseudo-samples after the probability integral transform. The pseudo-samples in the copula space are used as the input data for the quantum approach, and after the quantum approach generates new data in the copula space, the estimated inverse transformation $\hat{{F}_j}^{-1}$, $j=1,\dots,n$ is applied for study in the original return space. 

\section*{Quantum Formulation}
A general quantum state on $k$ qubits can be defined as follows:
\begin{align}
    |\Psi\rangle=\sum_{i=0}^{2^k-1}c_i |i\rangle,
\end{align}
where $|i\rangle$ ($i\in Z$) are known as ``computational basis states'', and $c_i$ are complex numbers with the condition that $\sum_i|c_i|^2=1$. In general, a quantum state can be prepared by the application of a quantum circuit to a set of qubits which are all initialized to the 0 state. Measurement on the qubits after the state is prepared is equivalent to sampling from a random number generator, where the probability of obtaining the random number $i$ is $|c_i|^2$. A parameterized quantum circuit can be optimized to learn the joint distribution of the variables in a given dataset. After training, the quantum circuit can be executed the desired number of times to produce samples for downstream applications. 

\begin{figure}[t]
\centering
 \includegraphics[width=0.48\textwidth]{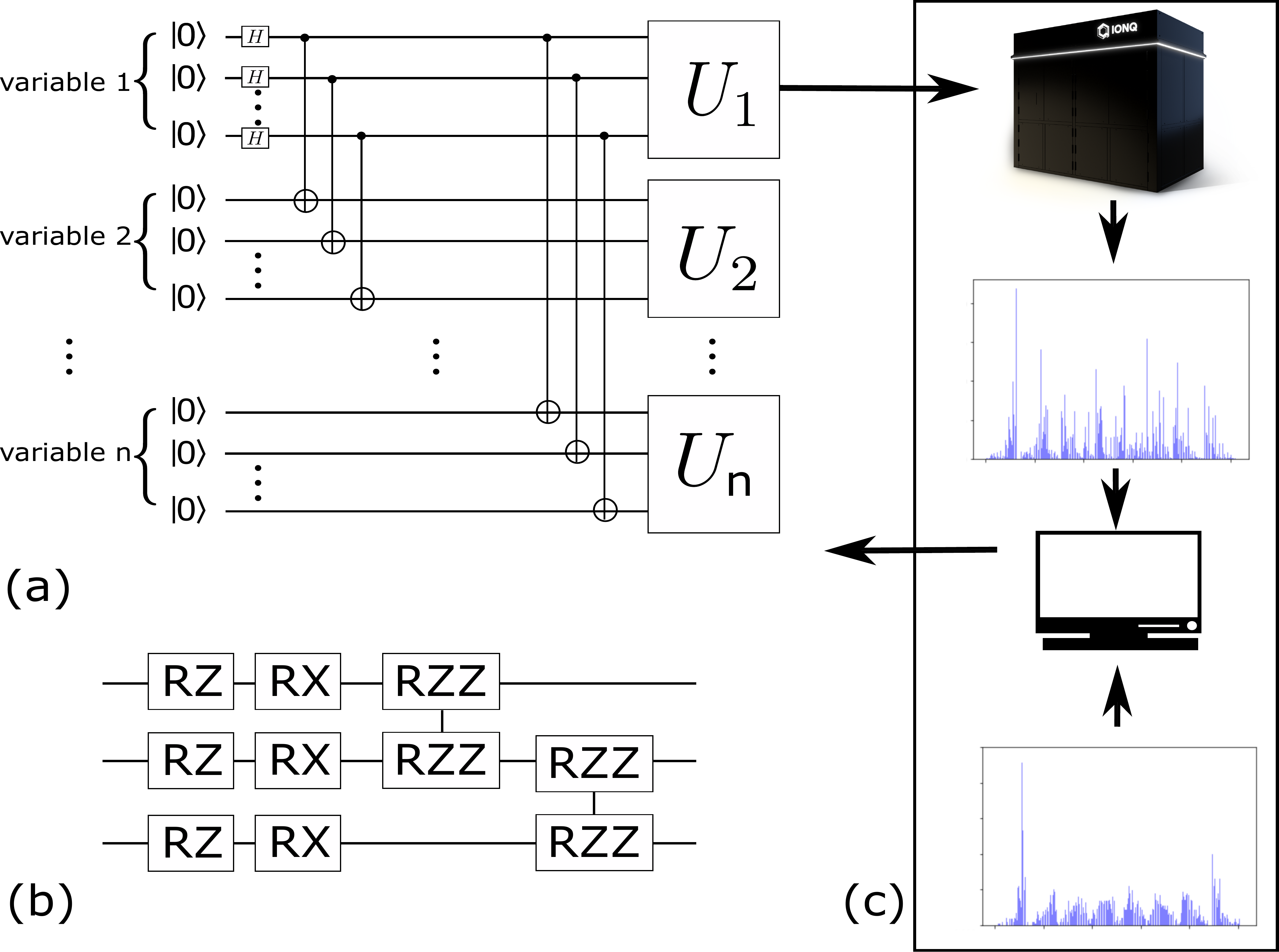}
 \caption{The hybrid training loop for the Quantum Circuit Born Machine (QCBM) framework: (a) The parametric ansatz used to model an $n$-variable copula. (b) The structure of each layer used to construct $U_i$. Each $U_i$ constructed with such layers is used to address their corresponding variable. (c) In each iteration of the optimization loop, the ansatz with a given set of parameters is executed on the quantum computer repeatedly to collect a statistical estimation of the output states, which is represented here as a histogram. It is then compared with a target histogram. The difference, quantified by the KL-divergence is used by a classical optimizer to drive the optimization. The optimization loop repeats until convergence.   }
 \label{fig:overall_circuit}
\end{figure}

The previous work \cite{zhu2021generative} proposed a parametric quantum circuit ansatz that prepares a quantum state corresponding to a discretized copula distribution. This ansatz was used to model the correlation between two variables by optimizing the circuit parameters based on the dataset consisting of the returns of two individual stocks. In this study we train a generalized version of the ansatz which can handle an arbitrary number of variables. 
The ansatz is shown in Fig.~\ref{fig:overall_circuit}. To model $n$-variable copulas discretized to precision of $m$ bits per variable, we need $m\times n$ qubits. These qubits are divided evenly among $n$ registers, where each register corresponds to one of the variables. Maximally entangled states called Greenberger–Horne–Zeilinger (GHZ) states are then formed which consist of one qubit from each register. At this point, the reduced density matrix of each register is an identity matrix, representing a standard uniform marginal. We then perform unitary transformations, denoted by $U_1,\dots,U_n$ in Fig.~\ref{fig:overall_circuit}, on each of the registers.
The unitary transformations $U_i$ are implemented via parameterized quantum circuits. In principle, there are infinitely many designs of circuits that can realize any specific $U_i$, as stated in \cite{sim2019expressibility}. In practice, the choices are often made to leverage native controls, as well as known symmetries of the target dataset. As a rule of thumb, implementations with deeper circuits will have more expressibility, which quantifies the capability of a parameterized quantum circuit to reach different points in the Hilbert space \cite{sim2019expressibility}. In our work, $U_i$ contains layers of the parametric circuit unit shown in Fig.~\ref{fig:overall_circuit}(b). Each of the gates has an individual parameter controlling the angle of the single- or two-qubit rotation operation. The parametric circuit units are optimized for hardware implementation. In particular, the arbitrary single-qubit rotations layer is implemented as a sequential application of a single-qubit rotation along the $z-$axis, $R_z(\theta)=\exp\bigg(-i\frac{1}{2}\theta\hat{\sigma}_z\bigg)$, and a single-qubit rotation along the $x-$axis, $R_x(\phi)=\exp\bigg(-i\frac{1}{2}\phi\hat{\sigma}_x\bigg)$, where $\hat{\sigma}_i$ stands for the Pauli matrices, and the three parameters $\theta$, $\phi$ and $\psi$ correspond to the rotation angles of the gates. According to the Euler decomposition, any arbitrary single-qubit rotation can be decomposed into a sequential application of $R_z(\theta)$, $R_x(\phi)$ and $R_z(\psi)$. We can commute the last $R_z(\psi)$ through the entangling gate $R_{zz}(\theta)=\exp(-i\theta\hat{\sigma}_z\otimes\hat{\sigma}_z)$ and merge into the first $R_z$ gate of the next layer. Here $\otimes$ is the tensor product, which indicates that the two Pauli matrices are applied on two different qubits. The $R_{zz}$ gate of arbitrary angle is decomposed into a controlled-not gate (CNOT), a $R_z$ gate, and another CNOT gate, for hardware implementation \cite{nielsen00}. 
The above construction of $U_i$ not only reduces the number of free parameters in optimization, but also maximizes the use of $R_z$ gates, which are implemented virtually, thereby being noise-free.
Similar to most of the popular universal quantum circuit ansatzes, with enough layers this structure is capable of representing any unitary transformation. 

We employ a hybrid circuit optimization framework for training. Within this framework, the parametric circuit ansatz creates a quantum state from which samples are generated according to the measurements on the qubits. The probability of getting a specific readout result is directly related to the amplitude of such state in the superposition. Because this is known as Born's rule, such a procedure is generally known as the Quantum Circuit Born Machine (QCBM) \cite{liu2018differentiable}. 
Within each iteration of the hybrid optimization loop, we evaluate the parametric ansatz for a specified number of repetitions to estimate the distribution of the generated samples. Then we compare the generated distribution against the distribution of the target data. The difference, quantified by a cost function, is then used to drive a optimizer to modify the variational parameters. For this work, we use 
the simultaneous perturbation stochastic approximation (SPSA) algorithm \cite{spall1998overview}
as the optimizer. A brief explanation of the SPSA algorithm is given in the supplementary materials.

To comprehensively appraise the training results of different implementations of copula modeling, we randomly split the data set by 80/20 into the training and testing set. After the model is trained, we can evaluate it with both an in-sample test (with the training set) and out-of-sample test (with the test set). These two tests benchmark not only how well the model trained, but also the utility of the trained model. 

To model the real-world data with quantum circuits, we apply the following conversion between real-valued pseudo-samples in the copula space and the binary strings represented by the qubits. First, depending on how many qubits are allocated to each variable, we digitize the real-valued pseudo-samples into binary strings. The binary strings of all variables are then concatenated into a single binary string. To elaborate, assume we use a $n$-variable ansatz, with $m$ qubits per variable. A sample in the copula space can be written as $(d_1,\dots,d_n)$ with $d_i\in [0,1)$. The copula-space sample is then converted into binary-valued samples $(b_1,\dots,b_n)$, where $b_i$ is the largest m-digit binary number $b_i=\overline{b_{i,0}...b_{i,m-1}}$ that $\frac{1}{2^m}\sum_{j=0}^{m-1}b_{i,j}2^j\leq d_i$. Here $b_{i,j}$ stands for the $j$-th digit of $b_i$. Then the digitized binary representation is combined into a single binary number $B=\sum_{i=0}^{n-1}b_i 2^{m\times i}$.  
Now this value can be exactly represented by measurement on the qubits. For example, assume we use two qubits for each variable. A pseudo-sample of two variables in the copula space $(d_1,d_2)=(0.735, 0.222)$ is first converted into $(b_1,b_2)=(10,00)$, and then combined into $1000$.
We use the distribution of binary strings converted from the training set as the target to train the quantum model.

The conversion of binary-valued measurement results of qubits to pseudo-samples is similar. We split each measurement on the qubits into $n$ $m$-digit binary numbers. We convert each binary number $b_i=\overline{b_{i,0}...b_{i,m-1}}$ back into a real-valued number $2^{-m}\sum_{j=0}^{m-1}b_{i,j}2^j= d_i{}'$, and then pad them with randomly generated numbers $\delta\in[0,\frac{1}{2^m})$.
In the prior example, a qubit readout $1000$ is first split into $(b_1,b_2)=(10,00)$. Then the pair is converted back to real number as $(d_1{}',d_2{}')=(0.5+\delta_1, 0+\delta_2)$, where $\delta_1$ and $\delta_2$ are independently drawn from $[0,0.25)$. Note the conversion into qubit representation is lossy due to the limited digits offered by the quantum model, while the conversion backwards is lossless.

To train the quantum model, as a common choice of cost function for a QCBM, we consider the Kullback–Leibler divergence (KL-Divergence) to capture the difference of two distributions \cite{zhu2019training}:
\begin{equation}
    D_{KL}(P,Q)=\sum_{x\in \chi}P(x)\log\bigg(\frac{P(x)}{Q(x)}\bigg).
\end{equation}
The KL-Divergence is asymmetric with respect to $P$ and $Q$. We set $P$ as  the distribution generated by the QCBM and set $Q$ as the target distribution in our hybrid training. To avoid numerical singularity, we use the clipped version of the KL-Divergence as \cite{zhu2019training}:
\begin{equation}
    D_{KL}(P,Q)=\sum_{x\in \chi}P(x)\log\bigg(\frac{P(x)}{\max(Q(x),\epsilon)}\bigg).
\end{equation}
The value of $\epsilon$ should be small enough so that it keeps the behavior of the KL-Divergence intact, yet large enough to prohibit numerical singularity. We heuristically set $\epsilon$ as $10^{-8}$.  

\section*{Results}

\begin{figure}[ht]
\centering
 \includegraphics[width=0.5\textwidth]{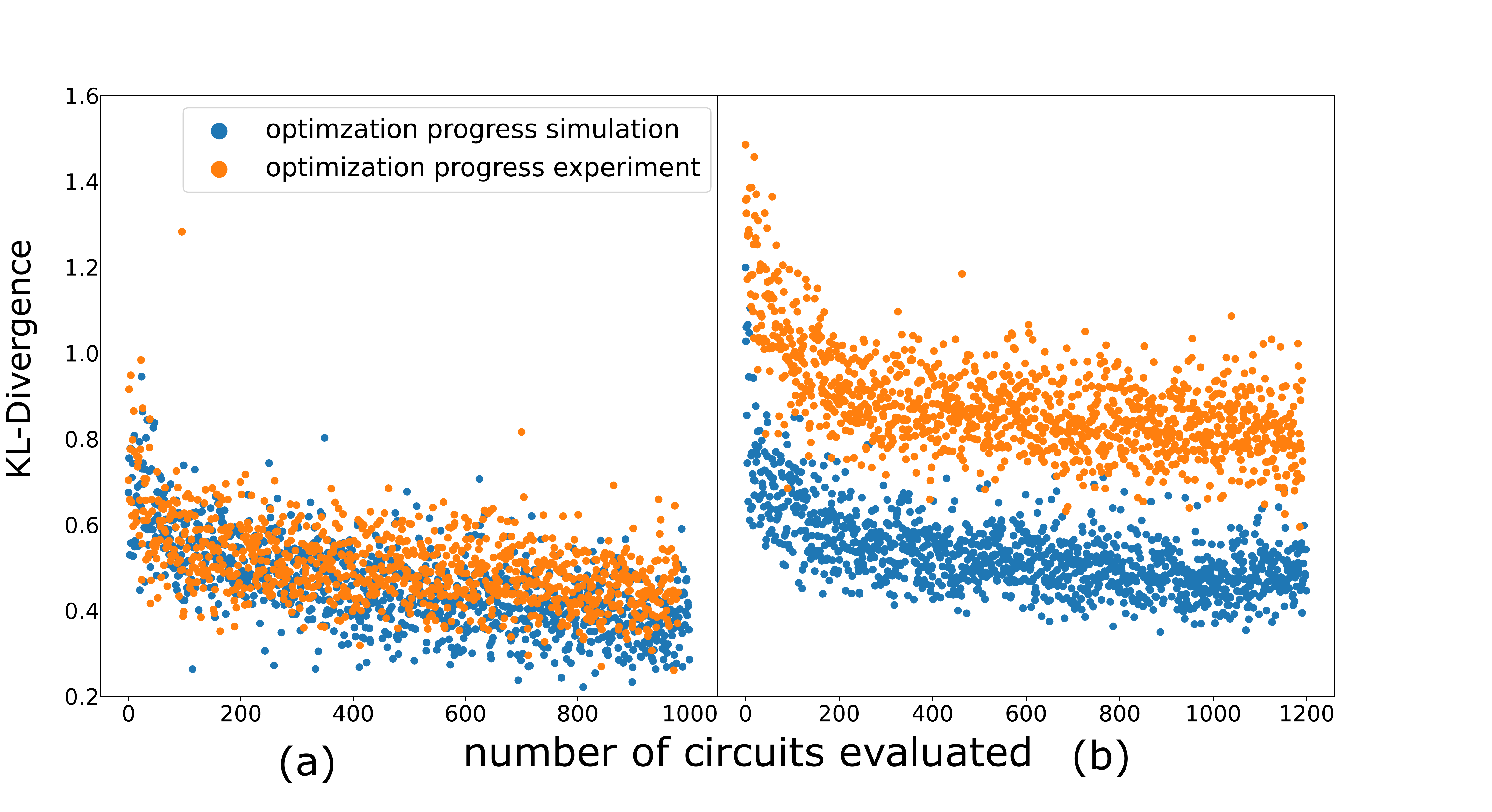}
 \caption{Optimization on a simulator vs. real quantum computer (experiment). (a) Hybrid optimization of a three-variable parametric ansatz. Each register of the ansatz includes two qubits. In both experiment and simulation, we use 500 SPSA optimization steps to optimize the ansatz. Each step involves executing two circuits to probe the gradient stochastically. The optimization in experiment and in simulation matches remarkably well. (b) Optimization of a four-variable parametric ansatz. Each register of the ansatz includes two qubits. In both experiment and simulation, we use 600 SPSA optimization steps to optimize the ansatz. Due to larger number of gates, the circuit execution suffers from experimental noise non-trivially which leads to a higher value of the final cost function. However, the optimization evolution in experiment and simulation qualitatively agree with each other.}
 \label{fig:training_progress}
\end{figure}

We present results from training the quantum models both on a simulator as well as trapped ion quantum hardware. 
The experimental demonstration was performed on the newest generation IonQ quantum processing unit (QPU). This system, as in previous IonQ QPUs \cite{wright2019benchmarking}, utilizes
trapped Ytterbium ions where two states in the ground hyperfine manifold are used as qubit states. These states are manipulated by illuminating individual ions with
pulses of 355 nm light that drive Raman transitions between the ground states defining the qubit. By configuring these pulses, arbitrary single qubit gates and Molmer-Sorenson type two-qubit gates can both be realized. Compared to its predecessors, this QPU features not only an order of magnitude better peak performance but also considerably better robustness in terms of gate fidelities. This allows for deep circuits with many shots to be run over a reasonable period of time. This increased data collection rate has made it possible to run hybrid optimization such as the one in this paper.

Fig.~\ref{fig:training_progress}(a) and (b) show two examples of training with hybrid quantum-classical optimizations involving 3 (DJI, VIX and N225) and 4 (DJI, VIX, N225 and RUT) variables, each with 2 qubits per variable. In both cases, the training on both the simulator and hardware converges, indicating that the training is practically scalable to higher than 2 dimensions studied in \cite{zhu2021generative}. In Fig.~\ref{fig:training_progress}(b), due to noise in the hardware, the experiment is unable to converge to as low of a minimum as the simulator. This effect is expected to be mitigated on future generations of hardware as the noise level becomes lower.

\begin{figure*}[ht]
\centering
 \includegraphics[width=\textwidth]{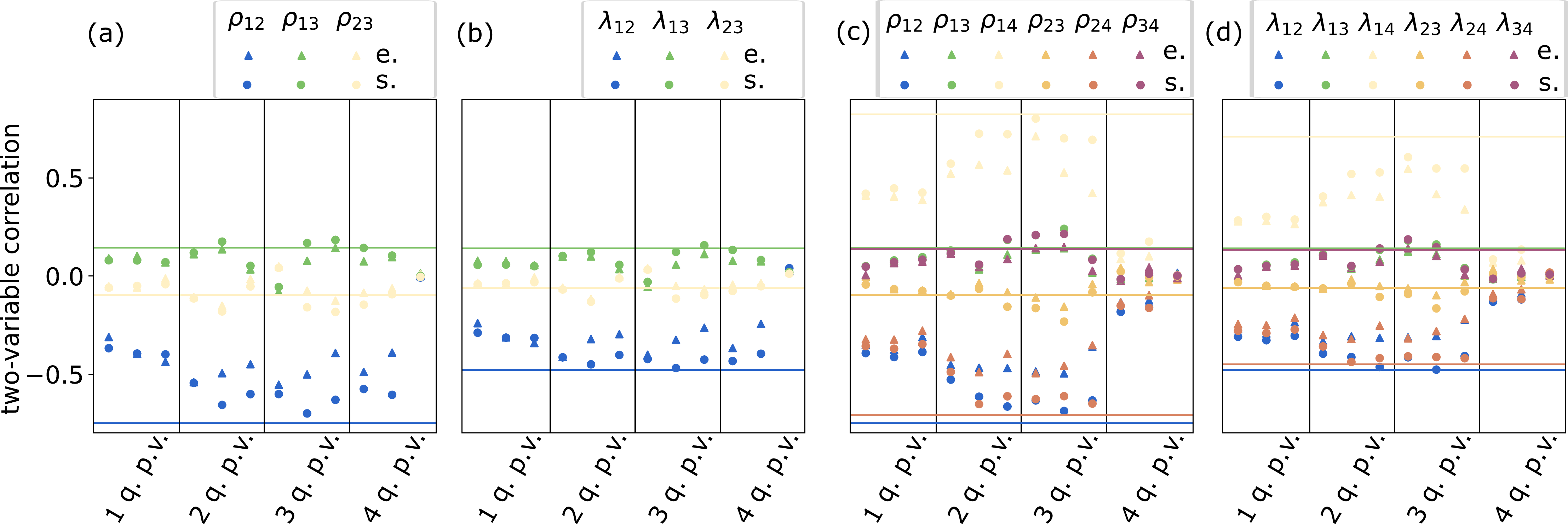}
 \caption{The two-variable correlators estimated using samples generated from experiment and samples generated from simulation, with different hyperparameters: number of qubits used for each variable; and number of layers (as shown in Fig.~\ref{fig:overall_circuit}) used to address each variable. Within each column indexed $i$-qubits per variable (abbreviated as ``q.p.v'' in the figure), results corresponding to 1-4 layers are ordered from left to right. The colored lines correspond to the ground-truth of the correlators shown with the same color. We use spheres to represent simulation results and triangle to represent results from the QPU. ``e.'' in the label stands for experiments. ``s.'' in the label stands for simulation. (a) Estimation of linear correlations $\rho$ using 3-variable copula of different configurations. (b) Estimation of upper tail dependence coefficients $\lambda$ using 3-variable copula of different configurations. (c) Estimation of linear correlations $\rho$ using 4-variable copula of different configurations. (d) Estimation of upper tail dependence coefficients $\lambda$ using 4-variable copula of different configurations. Note that $\rho_{ij}$ and $\lambda_{ij}$ represent the linear correlation and the upper tail dependence coefficient between the $i$-th and $j$-th variable, respectively. The error bars are omitted since they are of the size of the markers. The ground truths are based on the training data set, thereby representing an in-sample test.}
 \label{fig:correlator_plot}
\end{figure*}

We first discuss in-sample testing results. For each variable, we compute the basic statistics, including the mean, standard deviation, skewness, kurtosis, the 5-th percentile, the 25-th percentile, the 50-th percentile, the 75-th percentile and the 95-th percentile of the simulated samples, for both classical and quantum implementation. The results for the basic statistics from the classical and quantum methods all produce less than $0.5\%$ relative difference against the training data. Since all copulas have the standard uniform marginal, these tests have verified that our quantum ansatz in Fig.~\ref{fig:overall_circuit} has captured the standard uniform marginal for each variable, irregardless of the training processes. We omit concrete values for simplicity.

To examine the ability of our quantum model to learn the correlation between variables, we then report in-sample tests based on Pearson linear correlations $\rho$ and upper tail dependence coefficients $\lambda$. In particular, Pearson correlation mainly captures the linear relation of two random variables around the body of the associated distributions, whereas the upper tail dependence coefficient measures the co-movements of two random variables in the tails of the distribution \cite{caperaa1997nonparametric}. In financial applications, given a portfolio of multiple assets, the aggregated risk is mainly driven by the co-movements of the associated assets as the individual risks have been diversified away. When the market is volatile but remains far from capitulation, various aggregated risk measures can be chiefly captured by Pearson correlation. When the market crashes, as big losses are in the tails of the return distributions, tail dependence coefficients are representative measures. Specifically, that the tail dependence coefficient is nonzero indicates the tail of the distribution is heavier than normal distributions.   

Fig.~\ref{fig:correlator_plot} illustrates that correlations estimated from different configurations gradually approach the ground truth as we enhance the expressibility of the variational ansatz by including more qubits and more layers. The configuration with 3-qubits per variable, 3-layers of variational ansatz per variable yields the best performance. More complex models with 4-qubits per variable show inferior performance due to the degradation in the optimization processes. We believe that the close correlation results from both correlations by the quantum and the classical method underscore that the quantum method has well captured correlated movement of random variables.   

In the last column of Fig.~\ref{fig:correlator_plot} (c) and (d), the inferior performance of the models suggests that the optimization has failed completely. In this case, the number of basis states of the qubits are $2^{4 \times 4}>65,000$, which is the same as the number of different samples the same number of classical bits can generate. But the number of data points we supply to the training is only 4729, which is not enough to train a model with such a precision. In contrast, the next smallest models have $2^{4 \times 3}=4096$ basis states, by which the model are trained successfully. Consequently, we suggest that to train a QCBM, the number of qubits per variable should be chosen such that the number of basis states is less than the number of data points, so that the information included in the data is sufficient to train at such precision.

\section*{End-to-End Evaluation}

We perform out-of-sample validation for both the copula and the original return space for an end-to-end evaluation of the utility of the quantum framework. The latter are included in supplementary materials. 

After training, we check the accuracy of the estimated $95\%$ value at risk (VaR) and expected shortfall (ES) on the testing set for an equally-weighted portfolio \citep{mcneil2010quantitative,zhang2018review}. Financial institutions commonly calculate a spectrum of risk measures to quantify their risk exposure of their positions on a regular basis, and then reserve capitals accordingly for possible risk events due to internal needs or regulatory requirement \citep{goodhart2022holistic}. Among them, VaR is probably the most widely used risk measure. Given some confidence level $\alpha$, the VaR of a portfolio with loss $L$ at the confidence level $\alpha$ is given the smallest number $l$ such that the probability that the loss $L$ is no larger than $1-\alpha$. Formally, denoted by $F_L$ the cumulative distribution function of the loss, we have
\begin{equation}
    \mbox{VaR}_{\alpha} = \inf\{l\in \mathbb{R}: F_L(l)\geq \alpha\}. 
\end{equation}
VaR is simply a quantile of the loss distribution. ES is closely related to VaR and often reported with VaR as a coherent risk measure. Mathematically,
\begin{equation}
\mbox{ES}_{\alpha} = \frac{1}{1-\alpha}\int_\alpha^1 \mbox{VaR}_u(L)du.
\end{equation}
ES averages VaR over all levels $u\geq \alpha$ further into the tail of the loss distribution, and $\mbox{ES}_\alpha \geq \mbox{VaR}_\alpha$. Both are focused on the tail of the loss distribution as adverse co-movements of assets commonly lead to large loss on the tail. Hence, VaR and ES represent the further scrutiny of the modelled dependence structure on a portfolio besides Pearson correlations and tail dependence coefficients in Fig.~\ref{fig:correlator_plot}. As we only consider $\alpha=0.95$, the subscript is omitted for simplicity.

We report the ratio between the number of observed and expected failures for VaR, and report the ratio between the observed and expected severity for ES, where the failure is defined as the number of losses that are higher than the estimated VaR and severity is defined as the ratio between ES and VaR. Intuitively, 
the number of failures should be close to $(1-\alpha)$ of the tested samples for an accurate model, and the severity shows the extent to which ES captures tail risk that is not gauged by VaR. For ease of presentation, we call the former as the ratio of failures and the latter as the ratio of severity. By using ratios instead of raw values, we can more easily quantify the quality of the models in both copula space and return space in that observing near-one ratios in out-of-sample testing indicates the estimated VaR and ES are accurate. After training, before testing the decision rules of VaR and ES, VaR and ES are estimated on the data generated by the classical and the quantum approach, respectively. The accuracy of the estimated VaR and ES is evaluated by comparing the observed ratios with the expected ones on the testing data. In addition, all values are armed with the $95\%$ confidence interval characterizing the estimation error by bootstrapping.
\begin{figure*}[ht]
\centering
 \includegraphics[width=1\textwidth]{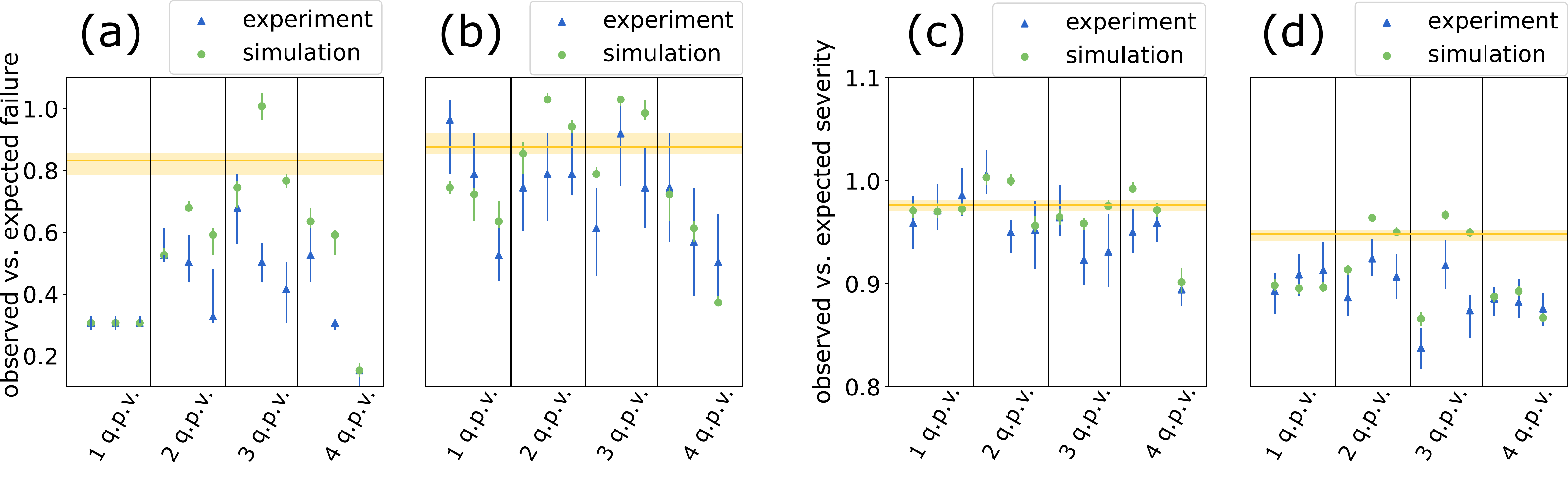}
 \caption{Out of sample test results for different QCBM models in the copula space, with different hyperparameters: number of qubits used for each variable; and number of layers (as shown in Fig.~\ref{fig:overall_circuit}) used to address each variable.. To perform the out-of-sample test, we split our data into a training set and a testing set, each contains $80\%$ and $20\%$ of the total samples exclusively. We use the training set to train our QCBM to generate samples to benchmark against the test data data set using two utility based metrics. The two metrics are based on an equally-weighted portfolio of the three and four indices, respectively. All values are armed a $95\%$ confidence interval. ``q.p.v.'' along the horizontal axes stands for ``qubits per variable''. Within each column indexed $i$-qubits per variable (abbreviated as ``q.p.v'' in the figure), results corresponding to 1-4 layers are ordered from left to right. (a) Ratio between observed and expected failures of the portfolio, aggregated with the 3-variable quantum model. (b) Ratio between observed and expected failures of the portfolio, aggregated with the 4-variable quantum model. (c) Ratio between observed and expected severity of the portfolio, aggregated with the 3-variable quantum model. (d) Ratio between observed and expected severity of the portfolio, aggregated with the 4-variable quantum model. The yellow line corresponds to the ratio obtained by the classical copula model. The lighter-yellow-colored region corresponds to the estimation error of the classical model results. Note that both the classical and the simulation approach estimate VaR and ES by 100K generated trials, while the experimental approach estimates VaR and ES by 5K generated trials.}
 \label{fig:OOS_test}
\end{figure*}

Fig.~\ref{fig:OOS_test} illustrates the two ratios estimated by different methods. The ratio of failures demonstrate higher variance as VaR is inherently more challenging to estimate than ES, while the ratio of severity is relatively stable as it takes average over losses on the tail in its calculation of ES. Both the classical and quantum methods are inclined to a conservative estimate for risk as the ratios are all not higher than one. Results from the simulator outperform the classical method in various configurations. Specifically, the optimal performance is roughly achieved around 3-qubits per variable and 1-layer structure for each $U_i$. The unimproved results for 4-qubits per variable, which is consistent with our observation in the correlator tests, is caused by the optimizer performance degradation. In the next section, we present a method to improve optimizer performance.  

%‘VaRLevel’ is for different level of confidence. If you want to just choose one, 0.95 is a good choice.  

%‘%ExpectedSeverity’ is the expected value of es/var, ‘ObservedSeverity’ is the corresponding observed value. The ratio of ‘ObservedSeverity’ / ‘ExpectedSeverity’ is not reported in the file.

%‘%Expected’ is the expected number of values that should exceed var, ‘Failures’ is the corresponding observed number. ‘Ratio’ is ‘Failures’/‘Expected’.

%Need definitions of quantities in %Figure 4.

%The two ratios are the suggested %y-values. 

%\begin{figure}[!t]
%\centering
%\includegraphics[width=1\linewidth]{Fig2.pdf}
%\caption{PV plant potential cyber-attack points: i) physical, ii) inverter controller and algorithm, iii) supply chain, iv) monitoring and diagnostics platform, and v) grid.}
%\label{fig:PVproattack}
%\end{figure}
\section*{Annealing Training Strategy}
\begin{figure}[ht]
\centering
 \includegraphics[width=0.4\textwidth]{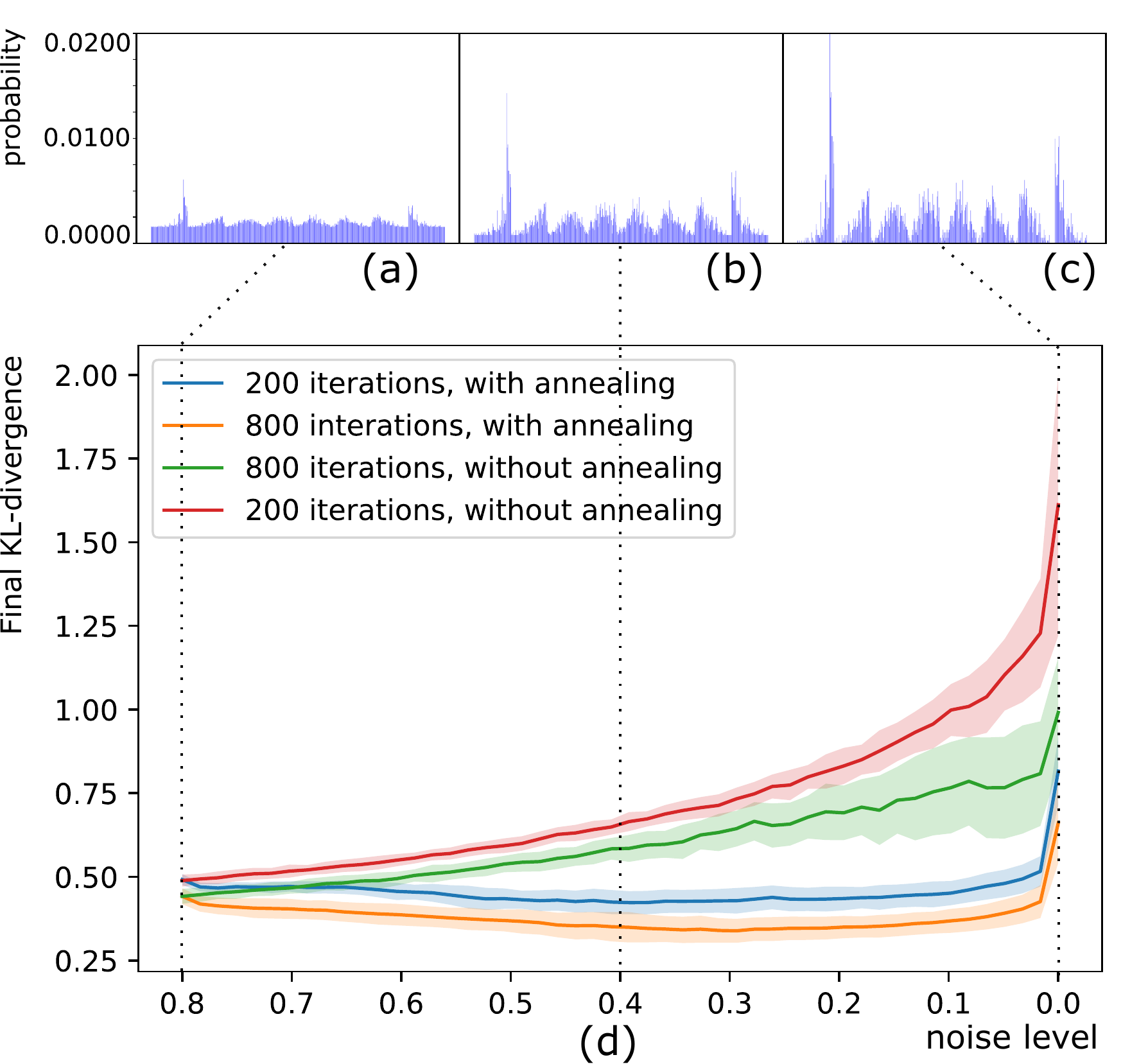}
 \caption{During annealing training, we start the training with a linear combination of the target distribution and with uniform distribution. Here, the initial target is a $20\%$ to $80\%$ mixture, which we refer to as uniform noise level $\eta=0.8$. We repeat ``ramping down'' steps, gradually decreasing the uniform noise level down to 0, at which point the target become the original target histogram. Within each ramping down cycle, we first decrease the noise level by the step size, then initialize the parameters of ansatz with the optimal parameters obtained from last ramping down cycle. Finally, we finish a cycle by repeating the optimization loops for a specified times. (a) Histogram of $20\%$ target distribution mixed with $80\%$ of uniform distribution. ($\eta=0.8$) (b) Histogram of $60\%$ target distribution mixed with $40\%$ of uniform distribution. ($\eta=0.4$) (c) Histogram of the original target distribution. ($\eta=0$) (d) Optimization results (as KL-divergence) obtained with and without annealing training. We use a step size of 0.02 ($\delta \eta=0.02$) for the ramping down. For annealing training, the optimization loops of each ramping down step is initialized to the final parameters from the last ramping down step. The results after each ramping down step is compared to the training with random initialization, but aimed at the same mixture of original target distribution and the uniform distribution. We explore two different number, 200 and 800, of optimization loops within each ramping down cycle. We observe that the annealing training, even with 4 times less iterations in each cycle, reaches a better final cost function. }
 \label{fig:annealing_training}
 
\end{figure}

Generally, optimization is a critical part of any hybrid quantum approach, including the QCBM. However, it is known that the optimization of quantum ansatzes can critically suffer from two issues: vanishing gradient and local minima of low utility. The vanishing gradient problem is also known as barren plateau \cite{mcclean2018barren}. As the expressiblity of a hybrid quantum ansatz grows, the gradient corresponding to the variational parameters in general decreases. Such a trend will eventually render the estimation of gradients impossible due to limited precision caused by the finite number of measurements possible. High-cost local minima are another common issue for both classical and quantum machine-learning \cite{gori1992problem,you2021exponentially,kawaguchi2020elimination}. The training or optimization procedure cannot guarantee the retrieval of global minimum. Instead, local minima are almost always obtained. Hence, even in the absence of vanishing gradient, effective strategies to converge to a useful local minimum via optimization is of great significance. 

We present an approach inspired by the adiabatic annealing process to address these two issues \cite{das2005quantum}. We call our approach 'annealing training'. The intuition is that by using an adaptive target (equivalently, cost function), we can first perform a relatively easy training, and then gradually increase the difficulty of the training. Assume $P_0=p_0(x)$ is the target distribution for our training. We define $P(\eta)=p(x,\eta)$ as a ``fuzzy'' target such that $p(x,\eta)=\eta u(x) + (1-\eta) p_0(x)$, where $u(x)$ is the uniform distribution of all the possible outputs $x$ of the qubits. We start the training by randomly initializing the parameters of the model and perform the desired number of optimization steps with $P(\eta_0)$ as the target. In principle, the annealing process should start with $\eta=1$ for the best performance. In practice, to save time in the annealing steps, $\eta$ can be initialized with a smaller value. This corresponds to starting annealing at a relatively low temperature, which may compromise the final training results depending on the specific cases. In our study, we heuristically start with $\eta_0=0.8$. We then repeat the ``ramping down'' steps, gradually decreasing $\eta$ down to 0, at which point the target becomes the original target $P_0$. Within each ramping down cycle, we first decrease $\eta$ by the chosen step size $\delta \eta$, then initialize the parameters of ansatz with the final parameters obtained from the last ramping down cycle. We finish a cycle by repeating the optimization steps for the desired number of iterations.

Analogous to the thermal annealing process, as long as the ramping down step size $\delta\eta$ is sufficiently small, there should be a high chance that the high-utility minima obtained from the last ramping down cycle are close to those of the current target. 

We compare the results from annealing training against those from standard training in Fig.~\ref{fig:annealing_training}. For each training methods, we perform the training with 800 and 200 iterations. With the same training method, more training iterations always generate better training results. But we observe that results obtained from annealing training outperform results obtained from standard training, even with 4-fold reduction in training iterations. We also observe that the standard deviation of training results, from stochastically choosing initial parameters, is smaller in annealing training. 
This is expected because through the annealing process, as the level of the induced uniform noise decreases, the minima will shift with the transformation of the cost-function landscape. But as long as such transformation is slow, the classical optimizer should drive the parameters to follow the shift of the minima. This also explains why annealing training can reach better results with fewer iterations. Because the parameters always start near local minima at each annealing step.  
Admittedly, repeating the annealing steps induces nontrivial overhead compared against the standard training. However, we expect such overheads to stay constant as the complexity of the model grows. Moreover, the overhead is typically acceptable if it leads to difference between failure and success of the training.

\section*{Conclusion and Outlook}
Modeling dependence of multiple variables is an essential part of understanding the world since it reflects the complicated interaction among moving parts. Given a system with a number of random variables, the mathematical formulation called copula can be used for capturing and modeling the underlying interdependencies. These interdependencies can then be leveraged to estimate the probabilities of extreme events of interest. Since its coinage in 1959, the copula has already been applied to many realms such as finance, engineering and medicine. 
%One of the most important resources of quantum computing is the quantum entanglement – the correlation among measurements of different qubits that has no classical correspondence. Taking advantage of the entanglement structure of the GHZ states, at scale our Quantum Circuit Born Machine (QCBM) based formula provides a promising way to perform copula modeling with higher training efficiency and provide better estimate to different values. 

Concurrently, with quantum entanglement, we can generate correlations among different qubits that have no classical correspondence. This provides a route to modeling copulas that can better estimate quantities relevant to risk management. To gain critical insight into the practicality of the quantum approach as the problem size scales up, we numerically and experimentally demonstrated modelling of 3- and 4- variable copulas of a variety of configurations, on real-world index data for stock markets. We showed effective training in both simulation and the latest generation quantum computer from IonQ. We observed that the effectiveness of conventional optimization methods  decreases as the complexity of the parametric quantum models grows. This complexity, mirroring the expressibility or raw power of the parametric quantum model, usually has to grow to accommodate larger problem sizes. To address this challenge, we introduced a novel optimization technique inspired by annealing, which greatly enhances the efficiency of training. This method provides an opportunity to extend variational quantum algorithms into the regime where the problem size would previously limit the efficacy of conventional optimizers. For future studies, it is of practical interest to characterize the annealing training more comprehensively by applying it to different types of problems.

We have performed in-sample and out-of-sample tests to evaluate multiple aspects of our trained models. Different from the prior work \cite{zhu2021generative}, we presented an end-to-end test using twenty years of data for four representative stock market indices with diverse dependence structures tested on various metrics in risk aggregation applications including the value at risk (VaR) and the expected shortfall (ES). The use of four indices with different underlying dependence structures and various risk metrics has allowed us to fairly appraise the proposed quantum framework from both in- and out-of-sample tests. It has also enabled us to pinpoint the numerical challenges in scaling to high-dimensional problems with more complex dependence structures. 

To conclude, about 60 years after the introduction of copulas, quantum computing is opening up new opportunities towards modelling and leveraging of dependence concepts. In particular, quantum copulas provide an additional tool for institutional investors in multi-asset return modeling for better risk assessment. This study takes quantum modelling of copulas several steps closer towards practical deployment and real-world impact. In future work, it is desirable to further characterize quantum advantage by extending our study to various sets of real-world data with a variety of interdependencies.

\showmatmethods{} % Display the Materials and Methods section

\acknow{A.G., W.S., S.R.M. and B.N. would like to thank Dave Vernooy for supporting quantum research – computing, networking and sensing – at GE Research}

\showacknow{} % Display the acknowledgments section

% Bibliography
%\bibliography{bibliography}

\newpage
% \section*{Supplementary Materials}

% \subsection*{SPSA optimizer}

% \subsection*{Tables for single-variable statistics}

\section*{Supplementary Materials}

\section{Out-of-sample Test in Return Space}

\begin{figure*}[ht]
\centering
 \includegraphics[width=1\textwidth]{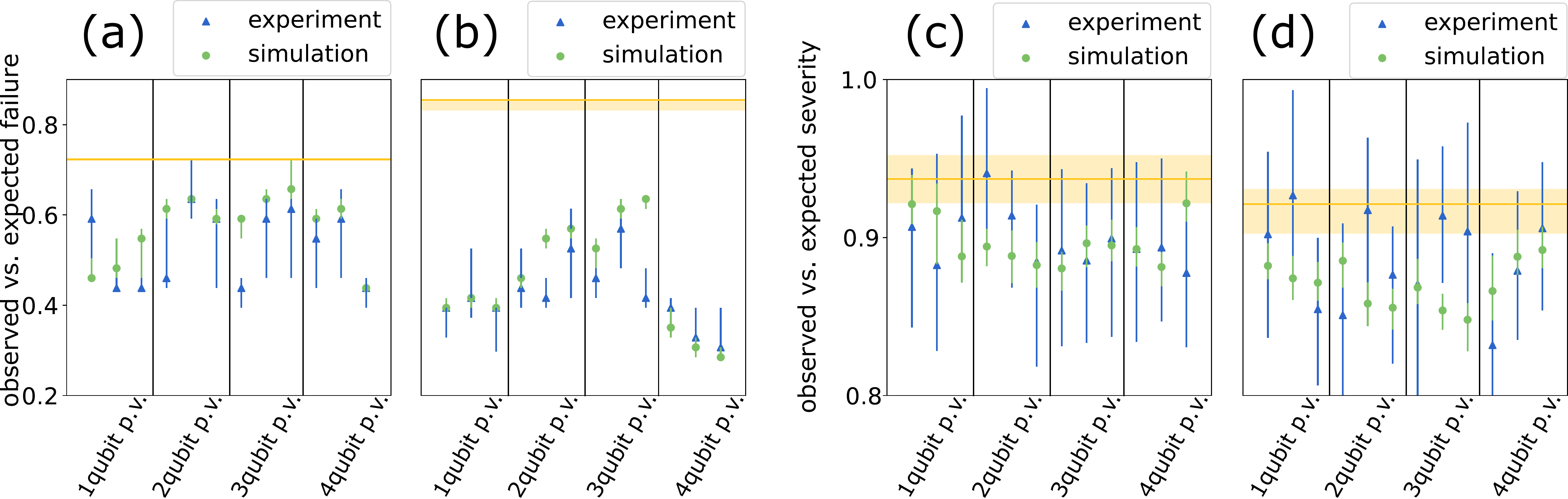}
 \caption{Out of sample test results for different QCBM models in the return space, with different hyper-parameters. See figure 6 in the main text for details of the out-of-sample test. (a) Ratio between observed and expected failures of the portfolio, aggregated with the 3-variable quantum model. (b) Ratio between observed and expected failures of the portfolio, aggregated with the 4-variable quantum model.(c) Ratio between observed and expected severity of the portfolio, aggregated with the 3-variable quantum model.  (d) Ratio between observed and expected severity of the portfolio, aggregated with the 4-variable quantum model. The yellow line corresponds to the ratio obtained by the classical copula model. The lighter-yellow-colored region corresponds to the estimation error of the classical model results. Note that both the classical and the simulation approach estimate VaR and ES by 100K generated trials, while the experiment approach estimates VaR and ES by 5K generated trials.}
 \label{fig:OOS_test_real_space}
\end{figure*}

In the main text, we presented the results of out-of-sample tests in copula space. The test is more representative of the quantum formula's accuracy because the model is trained in copula space. For the sake of completeness, we also perform the out-of-sample tests in the return space. The results are shown in Fig. \ref{fig:OOS_test_real_space}. 

Due to the inverse integral transformation which is nonlinear in nature, and neither convex nor concave, the error bars are amplified, compared with that in the copula space. The inverse transformation also amplifies the error at the tail portion of the distribution. Given the overlapping of confidence intervals and large variability, we can only state that the quantum models in experiment are comparable or relatively better than simulation.  However, after the transformation and calculation of severity, it is not easy to identify if the relatively better performance is inherent to the experiment or the effect of the transformation. This is a good question for a more theoretical paper on risk management. 

\section{Simultaneous Perturbation Stochastic Approximation (SPSA)}
We present a brief overview of the SPSA optimizer here for the completeness. A detailed explanation can be found in reference \cite{spall1998overview}.

The SPSA effectively uses gradient descent to optimize systems with many unknown parameters. But unlike the vanilla version of gradient descent that probes partial derivatives with respect to all the parameters, SPSA stochastically samples gradient along a single randomly selected direction using parameter-shift and adjusts the parameters accordingly in each step.  Because of the necessity of using parameter-shift rule for the estimation of derivatives in almost all hybrid quantum optimization procedures, as well as inevitable noise from sampling and hardware imperfection, such a stochastic approach provides great efficiency without compromising accuracy. 

Specifically, at each optimization step, the gradient $\hat{g}_{\hat{\Delta}}(\hat{\theta})$ along a random direction $\hat{\Delta}=(\delta_1,\dots,\delta_n)$ is obtained as $\hat{g}_{\hat{\Delta}i}(\hat{\theta})=\frac{f(\hat{\theta}+c\hat{\Delta})-f(\hat{\theta}-c\hat{\Delta})}{2c\delta_i}$. 
The optimizer than steps the parameter $\hat{\theta}$ along the direction of the probed stochastic gradient according to $\hat{\theta}=\hat{\theta}+a \hat{g}$.
Here $\delta_i$ are randomly drawn from $\pm1$, both $a$ and $c$ are hyperparameters that gradually decrease as the optimization proceeds.

In this work, we follow the recommendation of reference \cite{spall1998overview} to set both $a=c=0.1$ before hyperparameter tuning. After tuning, we found that setting $a=c=0.3$ yields optimal performance, largely due to various sources of noise.  

\section{Workflow}
\begin{figure*}[ht]
\centering
 \includegraphics[width=1\textwidth]{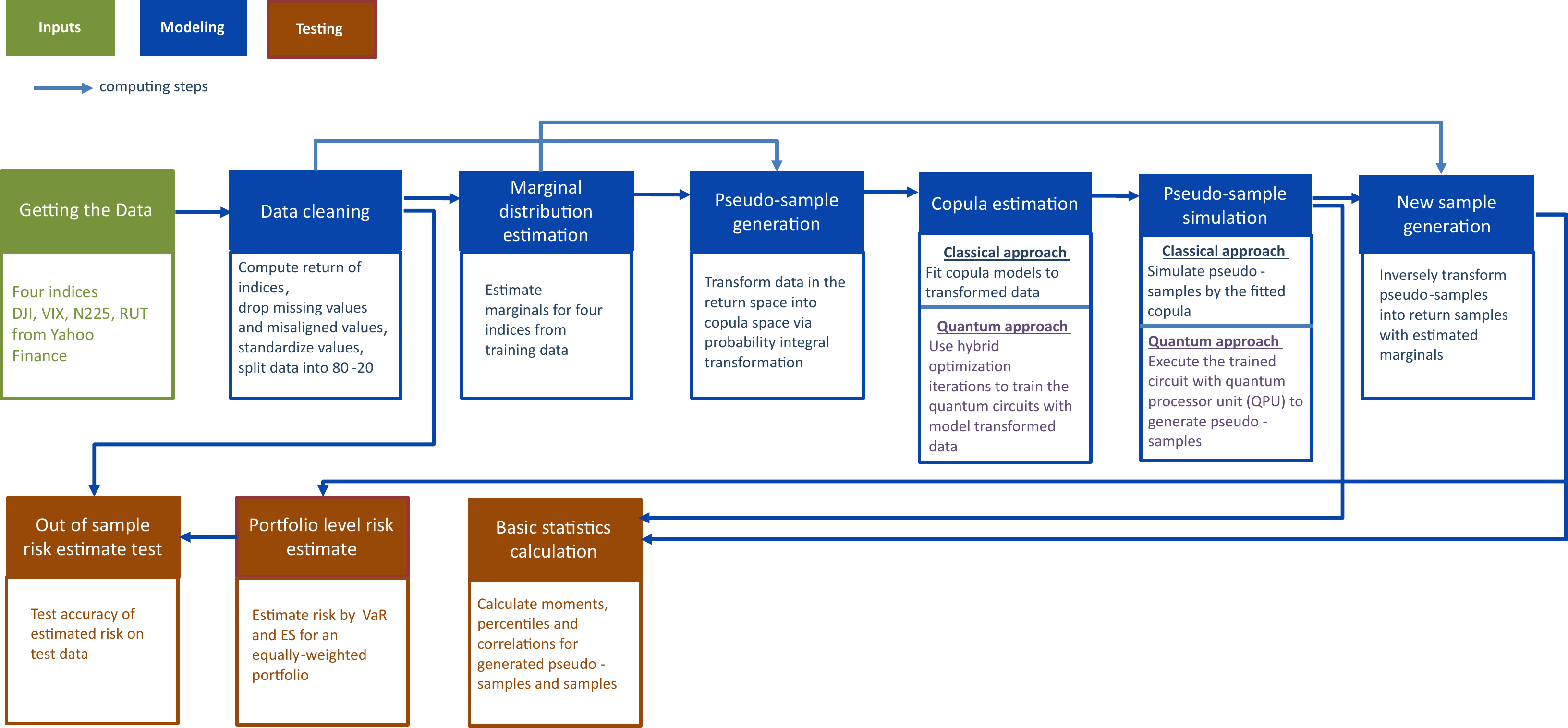}
 \caption{The flowchart illustrates the implementation of the end-to-end test for copula modeling, either with quantum or classical approach. The arrows corresponds to the direction of the data flow.}
 \label{fig:flow_chart}
\end{figure*}

In this section, we include a flowchart (Fig. \ref{fig:flow_chart}) that illustrates how the functional modules within our copula framework depend on each other, as well as the data flow. Note within our framework, all data flow are classical.

\end{document}